%% file: shggf.tex
\documentclass[10pt]{article}
\usepackage{colordvi}
\usepackage{epsfig}
\usepackage{axodraw}
\usepackage{epsfig}
\usepackage{graphicx}
\usepackage{rotate}
\usepackage{latexsym}
\usepackage{amssymb}
\usepackage{amsmath}
\usepackage[figuresright]{rotating}
\newcommand\pubnumber{}
\newcommand\pubdate{July 10, 2007}

\def\csumb{Dipartimento di Fisica Teorica, Universit\`a di Torino, Italy\\
INFN, Sezione di Torino, Italy}
\def\Title#1{\begin{center} {\Large\bf #1 } \end{center}}

\def\Author#1{\begin{center}{ \sc #1} \end{center}}
\def\Address#1{\begin{center}{ \it #1} \end{center}}

\newcommand\pubblock{\rightline{\begin{tabular}{l} \pubnumber\\
         \pubdate\\ \end{tabular}}}
\newenvironment{Abstract}{\begin{quotation}  }{\end{quotation}}

\def\Acknowledgments{\bigskip  \bigskip \begin{center}
          \large\bf Acknowledgments\end{center}}
\def\email#1{\footnote{#1}}
\makeatletter
\def\section{\@startsection{section}{0}{\z@}{5.5ex plus .5ex minus
 1.5ex}{2.3ex plus .2ex}{\large\bf}}
\def\subsection{\@startsection{subsection}{1}{\z@}{3.5ex plus .5ex minus
 1.5ex}{1.3ex plus .2ex}{\normalsize\bf}}
\def\subsubsection{\@startsection{subsubsection}{2}{\z@}{-3.5ex plus
-1ex minus  -.2ex}{2.3ex plus .2ex}{\normalsize\sl}}
\renewcommand{\@makecaption}[2]{%
   \vskip 10pt
   \setbox\@tempboxa\hbox{\small #1: #2}
   \ifdim \wd\@tempboxa >\hsize     
       \small #1: #2\par          
     \else                        
       \hbox to\hsize{\hfil\box\@tempboxa\hfil}
   \fi}
 \def\citenum#1{{\def\@cite##1##2{##1}\cite{#1}}}
\def\citea#1{\@cite{#1}{}}
\newcount\@tempcntc
\def\@citex[#1]#2{\if@filesw\immediate\write\@auxout{\string\citation{#2}}\fi
  \@tempcnta\z@\@tempcntb\m@ne\def\@citea{}\@cite{\@for\@citeb:=#2\do
    {\@ifundefined
       {b@\@citeb}{\@citeo\@tempcntb\m@ne\@citea\def\@citea{,}{\bf }\@warning
       {Citation `\@citeb' on page \thepage \space undefined}}%
    {\setbox\z@\hbox{\global\@tempcntc0\csname b@\@citeb\endcsname\relax}%
     \ifnum\@tempcntc=\z@ \@citeo\@tempcntb\m@ne
       \@citea\def\@citea{,}\hbox{\csname b@\@citeb\endcsname}%
     \else
      \advance\@tempcntb\@ne
      \ifnum\@tempcntb=\@tempcntc
      \else\advance\@tempcntb\m@ne\@citeo
      \@tempcnta\@tempcntc\@tempcntb\@tempcntc\fi\fi}}\@citeo}{#1}}
\def\@citeo{\ifnum\@tempcnta>\@tempcntb\else\@citea\def\@citea{,}%
  \ifnum\@tempcnta=\@tempcntb\the\@tempcnta\else
  {\advance\@tempcnta\@ne\ifnum\@tempcnta=\@tempcntb \else\def\@citea{--}\fi
    \advance\@tempcnta\m@ne\the\@tempcnta\@citea\the\@tempcntb}\fi\fi}
\makeatother
\input sbookp_rosetta.tex

\begin{document}
\begin{titlepage}
\pubblock
\vfill
\def\thefootnote{\fnsymbol{footnote}}
\Title{Complete Two-Loop Corrections to $H \to \gamma \gamma$ \footnote[4]{Work supported by MIUR under contract
2001023713$\_$006 and by  the European Community's Marie-Curie Research 
Training Network under contract MRTN-CT-2006-035505
`Tools and Precision Calculations for Physics Discoveries at Colliders'.}}
\vfill
\Author{Giampiero Passarino\email{giampiero@to.infn.it}, \hspace{0.1cm}
Christian Sturm\email{sturm@to.infn.it} \hspace{0.1cm} {\rm and} \hspace{0.1cm}
Sandro Uccirati\email{uccirati@to.infn.it}}
\Address{\csumb}
\vfill
\begin{Abstract}
\noindent 
In this paper the complete two-loop corrections to the Higgs-boson
decay, $H \to \gamma \gamma$, are presented. The evaluations of both QCD and
electroweak corrections are based on a numerical approach. The results
cover all kinematical regions, including the $WW\,$ normal-threshold,
by introducing complex masses in the relevant (gauge-invariant) parts of the 
LO and NLO amplitudes.
\end{Abstract}
\vfill
\begin{center}
\end{center}
\end{titlepage}
\def\thefootnote{\arabic{footnote}}
\setcounter{footnote}{0}
\small
\thispagestyle{empty}
%
\normalsize
\clearpage
\setcounter{page}{1}
\section{Introduction}
In the intermediate mass range of the Higgs-boson, its decay into photons is
of great phenomenological interest. At hadron colliders the decay $H \to
\gamma \gamma$ provides precious informations for the discovery in the
gluon-gluon production channel~\cite{Pieri:2006bm}.
An upgrade option at the ILC will allow for a high precision measurement
of the partial width into two photons~\cite{Moenig:2007py} with a quantitative
test for the existence of new charged particles.

The QCD corrections to $H \to \gamma \gamma$ have been computed in the past
and analytic results at next-to-leading order are available in 
Ref.~\cite{Spira:1995rr} and in Ref.~\cite{Harlander:2005rq}
(see also Ref.~\cite{Zheng:1990qa}). 
Electroweak two-loop corrections have been computed by suitable expansions 
of the two-loop Feynman diagrams~\cite{Degrassi:2005mc}. 
Master integrals for the two-loop light fermion contributions have been
analyzed in Ref.~\cite{Aglietti:2004ki} and two-loop light fermion contributions 
to Higgs production and decays in Ref.~\cite{Aglietti:2004nj}.

In our approach we have generated the full amplitude (up to two-loops
and including QCD)  in a completely independent way and we have used the
techniques of Ref.~\cite{Passarino:2001wv} to produce a numerical evaluation of
the partial width $\Gamma(H \to \gamma \gamma)$. Since we are not bound to rely on
expansion techniques, not even in the bosonic sector and in the top-bottom
one, we can produce results with very high accuracy for any value of the
Higgs-boson mass, taking into account the complete mass dependence of
the $W\,$-boson, $Z\,$-boson, Higgs-boson and top-quark. A consistent and
gauge-invariant treatment of unstable particles made it also possible
to produce very accurate results around the $WW\,$-threshold.
\section{Method of calculation and technical issues}
Our calculation builds upon the numerical approach
of Ref.~\cite{Passarino:2001wv} where two-loop, two and three point functions
have been investigated in the most general case. In this project we
have developed a set of routines which go from standard
$A_0,\,\dots\,,D_0$ functions up to diagrams needed for a two-loop $1 \to
2$ process. This new ensemble of programs will succeed to the
corresponding Library of {\tt TOPAZ0}~\cite{Montagna:1993ai}.
The whole collection of codes also uses the NAG-library~\cite{naglib}.

The generation as well as the manipulation of Feynman diagrams has been 
performed with the use of the FORM~\cite{FORM} code $\GS$~\cite{GraphShot}. 
Diagrams are generated, simplified and a FORTRAN interface is created. 
Furthermore, the code checks for the validity of the relevant Ward identities.
Renormalization is performed according to the scheme developed in
Ref.~\cite{Actis:2006rc}. In this paper, we shall follow the same notations
and conventions for two-loop diagrams as defined in
Ref.~\cite{Actis:2004bp}. In the following we will give a short outline
of the techniques used for the calculation. 

\begin{figure}[h]
$$
\raisebox{0.1cm}{
\scalebox{0.65}{
\begin{picture}(140,75)(0,0)
 \SetWidth{1.4}
 \Line(0,0)(40,0)                    \Text(10,5)[cb]{$-P$}
 \Line(128,-53)(100,-35)             \Text(138,-65)[cb]{$p_1$}
 \Line(128,53)(100,35)               \Text(138,57)[cb]{$p_2$}
 \CArc(100,-35)(70,90,150)           \Text(64,31)[cb]{$m_1$}
 \CArc(40,70)(70,270,330)            \Text(80,-2)[cb]{$m_2$}
 \Line(100,-35)(40,0)                \Text(70,-30)[cb]{$m_4$}
 \Line(100,-35)(100,35)              \Text(110,-3)[cb]{$m_3$}
                                     \Text(0,-25)[cb]{\Large $V^{\ssE}$}
                                     \Text(70,-70)[cb]{(a)}
\end{picture}
}
}
\!\!\!\!\!\Longleftrightarrow\;\;
\raisebox{0.1cm}{
\scalebox{0.65}{
\begin{picture}(140,75)(0,0)
 \SetWidth{1.4}
 \Line(0,0)(40,0)                    \Text(10,5)[cb]{$-P$}
 \Line(128,-53)(100,-35)             \Text(138,-65)[cb]{$p_1$}
 \Line(128,53)(100,35)               \Text(138,57)[cb]{$p_2$}
 \CArc(100,-35)(70,90,150)           \Text(64,31)[cb]{$m_2$}
 \CArc(40,70)(70,270,330)            \Text(80,-2)[cb]{$m_1$}
 \Line(100,-35)(40,0)                \Text(70,-30)[cb]{$m_4$}
 \Line(100,-35)(100,35)              \Text(110,-3)[cb]{$m_3$}
                                     \Text(70,-70)[cb]{(b)}
\end{picture}
}
}
\!\!\!\!\!\Longleftrightarrow\;\;
\raisebox{0.1cm}{
\scalebox{0.65}{
\begin{picture}(140,75)(0,0)
 \SetWidth{1.4}
 \Line(0,0)(40,0)                    \Text(10,5)[cb]{$p_2$}
 \Line(128,-53)(100,-35)             \Text(138,-65)[cb]{$p_1$}
 \Line(128,53)(100,35)               \Text(138,57)[cb]{$-P$}
 \CArc(100,-35)(70,90,150)           \Text(64,31)[cb]{$m_1$}
 \CArc(40,70)(70,270,330)            \Text(80,-2)[cb]{$m_2$}
 \Line(100,-35)(40,0)                \Text(70,-30)[cb]{$m_3$}
 \Line(100,-35)(100,35)              \Text(110,-3)[cb]{$m_4$}
                                     \Text(70,-70)[cb]{(c)}
\end{picture}
}
}
\!\!\!\!\!\Longleftrightarrow\;\;
\raisebox{0.1cm}{
\scalebox{0.65}{
\begin{picture}(140,75)(0,0)
 \SetWidth{1.4}
 \Line(0,0)(40,0)                    \Text(10,5)[cb]{$p_2$}
 \Line(128,-53)(100,-35)             \Text(138,-65)[cb]{$p_1$}
 \Line(128,53)(100,35)               \Text(138,57)[cb]{$-P$}
 \CArc(100,-35)(70,90,150)           \Text(64,31)[cb]{$m_2$}
 \CArc(40,70)(70,270,330)            \Text(80,-2)[cb]{$m_1$}
 \Line(100,-35)(40,0)                \Text(70,-30)[cb]{$m_3$}
 \Line(100,-35)(100,35)              \Text(110,-3)[cb]{$m_4$}
                                     \Text(70,-70)[cb]{(d)}
\end{picture}
}
}
$$
\vspace{0.7cm}
\caption[]{\label{Sym}Symmetries of the $V^{\ssE}$-family: The first
  diagram represents the $V^{\ssE}$-family (a). Its integral remains
  unchanged by exchanging $m_1\leftrightarrow m_2$ (b) as well as if one
  interchanges $m_3 \leftrightarrow m_4$ and $p_2 \leftrightarrow -P$
  simultaneously (c). The last diagram (d) is a combination of the first
  (b) and the second (c) symmetry. One can also perform a total
  reflection of all external momenta, which is not shown in the figure
  and leaves the integral also unchanged.}
\end{figure}
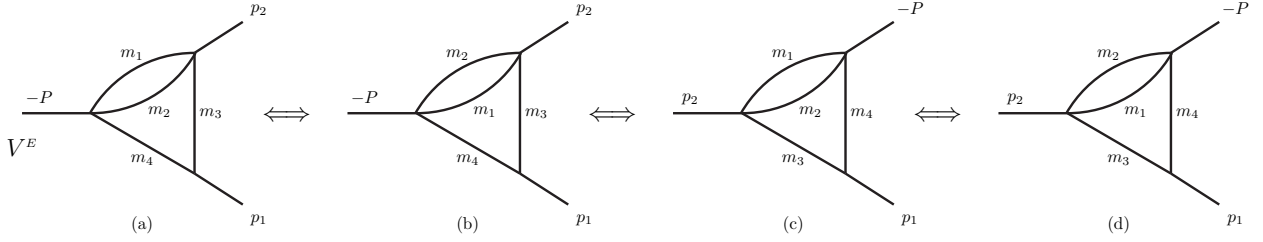

Before evaluating the two-loop integrals arising after generating the
Feynman diagrams, two main simplifications are done recursively.
At first, reducible scalar products are removed and secondly, the
symmetries of the diagrams are taken into account. The integrals are then
assigned to scalar-, vector- and tensor-type integrals, according to the
number of irreducible scalar products in the numerator and form-factors
are introduced. The cancellation of scalar products is performed by
expressing the scalar products in the numerator in terms of their
associated propagators. This procedure can lead to removing lines in a
diagram, so that each diagram produces a set of daughter-families with
at least one line less.  Apart from the reduction of scalar products,
the consideration of the symmetries of a given diagram is important in
order to reduce the number of integrals, which will be evaluated
numerically at the end of the calculation. A simple example, showing the
exploitation of the symmetries, is given in \fig{Sym} for a scalar
diagram.
\vspace{-0cm}

We now discuss briefly the extraction of collinear logarithms from Feynman 
diagrams. It is worth noting that the amplitude for $H \to \gamma \gamma$ is
collinear-free and one could adopt the approach where all light fermions are
massless, then collinear behavior of single components is controlled in 
dimensional regularization and collinear poles cancel in the total. We prefer 
another approach where collinear singularities are controlled by 
light fermion masses. Although the total amplitude is collinear-free,
our procedure of  reduction $\,\otimes\,$ symmetrization introduces a sum 
of several terms, of which some are divergent. Of course, we check that all 
logarithms of collinear origin cancel and, as a matter of fact, they cancel 
family by family of diagrams.

To be precise, we need some universal representation for the coefficient 
of the collinear logarithms, which allows us to show their analytical 
cancellation, and a method to compute the remaining collinear-free parts.
The first task is achieved by introducing integrals of one-loop functions 
and using their well-known properties to make the cancellation explicit. 
Using the techniques of Ref.~\cite{Passarino:2001wv} the collinear finite 
contribution is first written in terms of smooth integrands and then 
evaluated numerically; an example is shown in \fig{CollE}.
\vspace{-0.5cm}
\begin{figure}[h]
$$
\raisebox{0.1cm}{
\begin{picture}(140,75)(0,0)
 \SetWidth{1.2}
 \Line(0,0)(40,0)                    \Text(10,5)[cb]{$-P$}
 \Photon(128,-53)(100,-35){2}{5}     \Text(138,-65)[cb]{$p_1$}
 \Line(128,53)(100,35)               \Text(138,57)[cb]{$p_2$}
 \CArc(100,-35)(70,90,150)           \Text(64,31)[cb]{$M_1$}
 \CArc(40,70)(70,270,330)            \Text(80,-2)[cb]{$M_2$}
 \DashLine(100,-35)(40,0){3}         \Text(70,-30)[cb]{$m$}
 \DashLine(100,-35)(100,35){3}       \Text(110,-3)[cb]{$m$}
\end{picture}
}
\!\!\!\!\!
=\quad
-\,\,
\frac{1}{2}\,L_m^2\,\,
+\,\,
L_m\,
\intsx{x}\!
\quad
\raisebox{0.1cm}{
\begin{picture}(100,50)(0,0)
 \SetWidth{1.2}
 \Line(0,0)(30,0)                     \Text(10,5)[cb]{$-P$}
 \Line(70,0)(100,0)                   \Text(90,5)[cb]{$p_2$}
 \CArc(50,0)(20,0,360)                \Text(52,24)[cb]{$M_1$}
                                      \Text(52,-33)[cb]{$M_2$}
\Photon(5,-20)(30,0){2}{5}            \Text(8,-33)[cb]{$(1\!-\!x)\,p_1$}
\Photon(70,0)(95,-20){2}{5}           \Text(97,-33)[cb]{$x\,p_1$}
\end{picture}
}
\quad
+\,\,
\hbox{coll. finite}
$$
\vspace{1.4cm}
\caption[]{\label{CollE}Example of a collinear-divergent two-loop vertex
diagram. Dashed lines represent particles with a small mass $m$ and the 
wavy (external) line is massless. We introduced $L_m = \ln(m^2/|P^2|)$.}
\end{figure}
\section{Conceptual issues}
We will now apply our formalism to the computation of the amplitude for 
$H(-P) + \gamma(p_1) + \gamma(p_2) \to 0$, ($P= p_1+p_2$) which will be 
written as
\bq
{\cal A}^{\mu\nu}(H\to \gamma\gamma) =
\frac{g^3}{16\,\pi^2}\,\stws\,\Bigl[
F_{\ssD}\,\delta^{\mu\nu} + \sum_{i,j=1}^{2}\,
F^{(ij)}_{\ssP}\,p^{\mu}_i\,p^{\nu}_j + F_{\ep}\,\ep(\mu,\nu,p_1,p_2) \Bigr].
\label{basAmp}
\eq
The form factor $F_{\ep}$ is absent at $\ord{g^3}$ and only arises at
$\ord{g^5}$ but for a decay width with accuracy $\ord{g^8}$ (which
includes one-loop $\,\otimes\,$ two-loop) its contribution is again zero.
Bose symmetry and Ward identities (doubly-contracted, simply-contracted but 
with physical sources, simply-contracted with off-shell photons and 
unphysical sources) allow us to write the amplitude as
\bq
{\cal A} = 
\frac{g^3}{16\,\pi^2}\,\stws\,e_{\mu}(p_1)\,\Bigl[ 
F_{\ssD}\,\delta^{\mu\nu} + F_{\ssP}\,p^{\mu}_2\,p^{\nu}_1\Bigr]\,
e_{\nu}(p_2)=
\frac{g^3\,\stws}{16\,\pi^2}\,A,
\qquad
F_{\ssP} \equiv F^{(21)}_{\ssP},
\eq
where the form factors are expanded up to two-loops,
\bq
F_{i} = F_{i}^{(1)} + \frac{g^2}{16\,\pi^2}\,F_{i}^{(2)},
\quad
i=D,P,
\qquad\quad
A = A^{(1)} + \frac{g^2}{16\,\pi^2}\,A^{(2)}.
\eq
To proceed we need to include the relations between renormalized 
masses (small letters) and experimental, on-shell, ones (capital letters).
Finite renormalization is then completed by introducing external 
wave-function factors ($Z_{\ssH}^{-1/2}\,Z_{\ssA}^{-1}$) and the 
renormalization of the coupling constants. All needed relations are collected 
in \eqn{Hwfr}.
\bqa
&&
m_{\ssB}^2= M_{\ssB}^2\,\Bigl[
1 + \frac{\gf\,\mws}{2\,\sqrt{2}\,\pi^2}\,\Reb\,\Sigma^{(1)}_{\ssB\ssB}(M_{\ssB}^2)
\Bigr]
\quad
B= W,H,
\qquad\quad
m_t^2= M_t^2\,\Bigl[
1 + \frac{\gf\,\mws}{\sqrt{2}\,\pi^2}\,\Reb\,\Sigma_t^{(1)}(M_t^2)
\Bigr]
\nl
&&
g^2\,\stws\,Z_{\ssA}^{-1}= 4\,\pi\,\alpha,
\qquad\quad
g\,Z^{-1/2}_{\ssH}= 
2\,(\sqrt{2}\,\gf\,\mws)^{1/2}\,\Bigl[
1 - \frac{\gf\,\mws}{4\,\sqrt{2}\,\pi^2}\,\Pi_{\ssH}(\mhs)\Bigr],
\nl
&&
\Pi_{\ssH}(s)= \frac{\mhs}{s-\mhs}\,\Reb\,\Bigl[
\Sigma^{(1)}_{\ssH\ssH}(s) - \Sigma^{(1)}_{\ssH\ssH}(\mhs)\Bigr] -
\Reb\,\Sigma^{(1)}_{\ssWW}(\mws) +
\Sigma^{(1)}_{\ssWW\,;\,\ssC}(0),
\label{Hwfr}
\eqa
where $\Sigma^{(1)}_{\ssWW}$, $\Sigma^{(1)}_{\ssH\ssH}$ and 
$\Sigma_t^{(1)}$ are respectively the Higgs, $W$ and top quark one-loop 
self-energies as defined in section 5.3 of the second paper of 
Ref.~\cite{Actis:2006rc}; furthermore,
\bq
\Sigma^{(1)}_{\ssWW\,;\,\ssC}(0) = \Sigma^{(1)}_{\ssWW}(0) + \delta_{\ssG},
\qquad\quad
\delta_{\ssG} = 6 + \frac{7 - 4\,\stws}{2\,\stws}\,\ln\ctws,
\qquad\quad
\ctws= \frac{\mws}{\mzs}.
\eq
The symbols $M_t, \mw, \mz, \gf$ and  $\alpha$ denote the mass of the
top-quark, the $W\,$-boson, the $Z\,$-boson as well as the Fermi-coupling
constant and the fine structure constant.
Collecting all the ingredients we get the corresponding $S$-matrix completely 
written in terms of experimental data
\bqa
\label{tfinR}
{\cal A}_{\rm phys} 
&=& 
\Bigl(\sqrt{2}\,\gf\,\mws\Bigr)^{1/2}\,\frac{\alpha}{2\,\pi}\,
\Biggl\{\,\,
  A^{(1)}_{\,\rm ex}\,\,
+ \,\,\,\frac{\gf\,\mws}{2\,\sqrt{2}\,\pi^2}\,
  \biggl[\,\,\,
    A^{(2)}_{\,\rm ex}\,
  - \,\frac{1}{2}\,A^{(1)}_{\,\rm ex}\,\Pi_{\ssH}(\mhs)\,
\\
&+& 
    \,\,\mhs\,
    \frac{\partial A^{(1)}}{\partial\,m^2_{\ssH}}\bmid_{\rm ex}\!
    \Reb\,\Sigma^{(1)}_{\ssH\ssH}(\mhs)\,\,
  + \,\,\mws\,\frac{\partial A^{(1)}}{\partial\,m^2_{\ssW}}\bmid_{\rm ex}\!
    \Reb\,\Sigma^{(1)}_{\ssW\ssW}(\mws)\,\,
  + \,\,2\,M_t^2\,
    \frac{\partial A^{(1)}}{\partial\,m^2_t}\bmid_{\rm ex}\!
    \Reb\,\Sigma_t^{(1)}(M_t^2)\,\,
  \biggr]\,
\Biggr\}.
\nn
\eqa
The subscript ``ex'' indicates that all masses are the experimental ones and
the mass-shell limit ($s \to \mhs$) is taken only after the inclusion
of finite renormalization. QCD corrections will appear in \eqn{Hwfr} and
\eqn{tfinR} multiplied by $\pi\alpha_s(\mh)/(\sqrt{2}\,\gf\mws)$.

In our calculation we prove the cancellation of the collinear logarithms and
then set the light fermion masses to zero; therefore, due to Yukawa couplings,
an imaginary part in $A^{(1)}$ arises only if $\mh > 2\,\mw$. For 
two-loop terms imaginary parts are always present even for massless fermions.
From \eqn{tfinR} the total amplitude for $H \to \gamma \gamma$ can be 
written symbolically as
${\cal A}^{\mu\nu}_{\rm phys} = {\cal A}^{\mu\nu}_1\,\otimes\,\lpar 1 + {\rm FR}\rpar + 
{\cal A}^{\mu\nu}_2$.
Finite renormalization (FR) amounts to expressing renormalized parameters in 
the one-loop amplitude in terms of data and in the insertion of the Higgs 
wave-function factor $Z_{\ssH}$ \'a la LSZ; both requires the notion of on-shell mass.
There are two sources of inconsistency in this approach: the Higgs-boson is an
unstable particle and this fact has a consequence which shows up at two-loops.
When we compute the doubly-contracted Ward identity for the full two-loop 
amplitude we obtain
\bq
p_{1\mu}\,p_{2\nu}\,{\cal A}^{\mu\nu}_{\rm phys} = \lpar \gf\,\mws\rpar^{3/2}\,
\alpha\,\Imb\,W\lpar \mh,\mw,\dots\rpar.
\label{vWI}
\eq
The analytical form of $W$ is known and the non-zero result comes from
the fact that the pure two-loop contribution to the Ward identity gives
$W$ while finite renormalization gives the real part $\Reb\,W$.
Therefore, the Ward identity is violated above the $WW\,$-threshold.  On
top of this problem we find a second unphysical feature: let us analyze
how the amplitude for $H \to \gamma \gamma$ behaves around a
normal-threshold, i.e. for $\mh = 2\,\mw, 2\,\mz, 2\,M_t$.  In
particular we are interested in the question of possible square-root or
logarithmic singularities. Even if present they are unphysical, although
integrable. Both problems can be solved by using complex masses as
discussed in subsection~\ref{subsec:complexmass}.
\subsection{Square-root singularities}
It is very simple to prove that derivatives (represented by a dot in
\eqn{dot}) of one-loop, two-point functions with equal masses ($m$) develop 
a square root singularity:
\vspace{-0.5cm}
\bq
\raisebox{0.1cm}{
\begin{picture}(85,40)(0,0)
 \SetWidth{1.2}
 \Line(0,0)(25,0)                     \Text(10,5)[cb]{$p^2$}
 \Line(55,0)(80,0)                   
 \CArc(40,0)(15,0,360)                \Text(40,18)[cb]{$m$}
                                      \Text(40,-26)[cb]{$m$}
 \CCirc(40,-15){2.5}{1}{1}
\end{picture}
}
=\;
{\dot B}_0(p^2,m,m)\;
=\;
-\,\,
\frac{1}{\beta}\,\ln\frac{\beta+1}{\beta-1},
\qquad\qquad\quad
\beta^2= 1 + \frac{4\,m^2}{p^2-i\,\ep}.
\vspace{0.5cm}
\label{dot}
\eq
The same argument can be repeated for all the one- and two-loop diagrams 
with any number of external legs where we can cut two and only two $m$-lines; 
normal-threshold will be a sub-(sub-$\,\dots\,$)leading singularity, but a 
$1/\beta$-behavior shows up only if the reduced sub-graph responsible for 
the singularity can be reduced to a ${\dot B}_0\,$-function. 
Therefore the only two-loop vertex giving rise to a $1/\beta$-divergent 
behavior is the one depicted in \fig{hasCd}.
For this diagram it is possible to find a representation where the singular 
part is completely written in terms of one-loop diagrams, as shown in the
figure. The remainder can be cast in a form suited for numerical integration.
\vspace{-0.5cm}
\begin{figure}[ht]
$$
\raisebox{0.1cm}{
\begin{picture}(140,75)(0,0)
 \SetWidth{1.2}
 \Line(0,0)(40,0)                   \Text(10,5)[cb]{$P^2$}
 \Line(128,-53)(100,-35)            \Text(138,-65)[cb]{$p_1^2$}
 \Line(128,53)(100,35)              \Text(138,57)[cb]{$p_2^2$}
 \CArc(70,-17.5)(15,0,360)   
 \Line(57,-10)(40,0)                \Text(46,-16)[cb]{$m$}
 \Line(100,-35)(83,-25)             \Text(89,-42)[cb]{$m$}
 \Line(100,35)(40,0)                \Text(68,23)[cb]{$m$}
 \Line(100,-35)(100,35)             \Text(110,-3)[cb]{$m$}
\end{picture}
}
\!\!
=
\quad
-
\quad
\raisebox{0.1cm}{
\begin{picture}(80,40)(0,0)
 \SetWidth{1.2}
 \Line(-2,0)(25,0)                     \Text(12,5)[cb]{${}_{-}m^2$}
 \Line(55,0)(80,0)                   
 \CArc(40,0)(15,0,360)
\end{picture}
}
\,\,\times\,\,
\raisebox{0.1cm}{
\begin{picture}(80,40)(0,0)
 \SetWidth{1.2}
 \Line(0,0)(25,0)                     \Text(10,5)[cb]{$P^2$}
 \Line(59,-20)(76,-30)                \Text(80,-43)[cb]{$p_1^2$}
 \Line(59,20)(76,30)                  \Text(80,35)[cb]{$p_2^2$}
 \Line(25,0)(59,20)                   \Text(40,15)[cb]{$m$}
 \Line(59,20)(59,-20)                 \Text(68,-3)[cb]{$m$}
 \Line(59,-20)(25,0)                  \Text(40,-22)[cb]{$m$}
 \CCirc(42,-10){2.5}{1}{1}
\end{picture}
}
\quad
+
\quad
\left(
\ba{l}
\hbox{reg. part}\\
\hbox{at } \beta=0
\ea
\right)
$$
\vspace{1.5cm}
\caption[]{\label{hasCd} Contraction of a $V^{\ssM}$ configuration
leading to a $\beta^{-1}$-behavior at the normal $m$-threshold.}
\end{figure}

In the decay $H \to \gamma \gamma$, the $1/\beta$-singularity 
($\beta^2 = 1-4\,\mws/\mhs$) arises from the two-loop diagram of \fig{hasCd}, 
from Higgs-boson wave-function factor (derivative of a B-function) and from
finite $W\,$-mass renormalization (derivative of a C-function). 
Our conclusion is that the unphysical $1/\beta$-behavior around some normal-threshold is induced by self-energy like insertion, a fact that is not 
surprising at all; those insertions, signaling the presence of an unstable 
particle, should not be there and complex poles should be used instead.
\subsection{Logarithmic singularities}
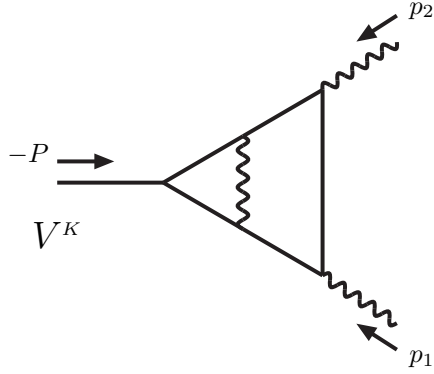
\begin{figure}[!ht]
\begin{center}
\begin{picture}(150,75)(0,0)
 \SetWidth{1.5}
 \Line(0,0)(40,0)         
 \Photon(128,-53)(100,-35){2}{5}  
 \Photon(128,53)(100,35){2}{5}    
 \Photon(70,-17.5)(70,17.5){2}{5}
 \Line(70,-17.5)(40,0)
 \Line(70,17.5)(40,0) 
 \Line(100,-35)(70,-17.5)
 \Line(100,-35)(100,35)
 \Line(100,35)(70,17.5)
 \LongArrow(0,8)(20,8)          \Text(-11,7)[cb]{$-P$}
 \LongArrow(128,-63)(114,-54)   \Text(138,-70)[cb]{$p_1$}
 \LongArrow(128,63)(114,54)     \Text(138,62)[cb]{$p_2$}
 \Text(0,-25)[cb]{\Large $V^{\ssK}$}
\end{picture}
\end{center}
\vspace{2cm}
\caption[]{\label{TLvertbca} The irreducible, scalar, two-loop vertex
diagram $V^{\ssK}$ with logarithmic divergency. Solid lines represent a
massive particle with mass $m$, whereas wavy lines correspond to
massless particles.}
\end{figure}

Let us consider the two-loop diagram of \fig{TLvertbca} with $P^2 = -s$ ($s > 0$). 
Writing the corresponding integral in parametric space we introduce the
quadratic forms 
\bq
\chi(x) = \lpar x - \frac{1}{2}\rpar^2 - \frac{1}{4}\,\beta^2, \qquad
\xi(x,y) = x \,\lpar x  - 1 \rpar\,y^2 + \frac{1}{4}\,\lpar 1 - \beta^2\rpar,
\qquad \beta^2 = 1 - \frac{4\,m^2}{s}
\eq
and obtain 
\bq
V^{\ssK} = \frac{2}{s^2}
           \int_0^1\,\frac{dx\,dy}{y\,\chi(x)}\,\Bigl[
\li{2}{1 - \frac{y\,\chi(x)}{\chi(x y)}} -
\li{2}{1 - \frac{y\,\chi(x)}{\xi(x,y)}}\Bigr].
\eq
Since we are interested in the behavior around $\beta \to 0$, we split 
$V^{\ssK}$ into a singular and regular part and find
\bq
V^{\ssK} = V^{\ssK}_{\rm sing} + V^{\ssK}_{\rm reg} 
         = \frac{2}{s^2}\int_0^1\,\frac{dx\,dy}{y\,\chi(x)}\,\Bigl[
\li{2}{1 - \frac{y\,\chi(x)}{\chi(x y)}} - \zeta(2) \Bigr]
	 +V^{\ssK}_{\rm reg}.
\eq
The singular part $V^{\ssK}_{\rm sing}$ will be written
as~\cite{Ferroglia:2002mz} 
\bq
V^{\ssK}_{\rm sing} = \frac{2}{s^2} \intfx{t}\,\frac{\ln t}{1-t}\,I(t),
\quad
I(t) =\int_0^1 dx\,dy\,\Bigl[ (1-t)\,\chi(x y) + t y \chi(x)\Bigr]^{-1} =
\int_0^1 dx\,dy\,\Bigl[ a\,\lpar x - X\rpar^2 + \lambda\Bigr]^{-1},
\label{VKsing}
\eq
where we have introduced the shorthands
\bq
a = \tau y, \qquad X = \frac{1}{2\,\tau}, \qquad
\lambda=  \frac{1}{4}\,\frac{t\,(1-t)}{\tau}\,\Bigl[ (1-y)^2 -
\beta^2\,\lpar y + T\rpar\,\lpar y + \frac{1}{T}\rpar\Bigr],
\eq
with $\tau = (1-t)\,y + t$ and $T = t/(1-t) > 0$. $I(t)$ can be split into 
two parts,
\bq
I(t)= B\lpar \frac{1}{2},\frac{1}{2}\rpar\,
\int_{y_{min}}^1\!\!\!dy\;a^{-1/2}\,\lambda^{-1/2} 
- \frac{1}{2}\,\sum_{i=1,2}\,
\int_0^1 dx\,dy\,
(-1)^i\,X_i\,x^{-1/2}\,\lpar a X^2_i + \lambda x\rpar^{-1},
\label{singsplit}
\eq
with $X_1 = - X$, $X_2 = 1 - X$ and B(x,y) is the Euler beta-function. 
While the second term of \eqn{singsplit} is regular for $\beta=0$, the
singularity of the first term follows from the fact that 
$\lambda \sim (1-y)^2$ for $\beta\to0$; however, we have a singular behavior only if 
$0 \le X \le 1$ which requires $y \ge y_{\rm min}=\max \{0\,,\,(t-1/2)/(t-1)\}$.
Since we are interested in the leading behavior for $\beta \to 0$, 
we can extend the integration domain in the first term to $[0,1]$, without
modifying the divergent behavior of the diagram.
The singular part is then given by
$$
I_{\rm sing}(t) = 2\,\pi\,\bigl[ t (1-t)\bigr]^{-1/2}\!
\intsx{y}\,y^{-1/2}\,\Bigl[ (1-y)^2 - \beta^2\,
\lpar 1 + T\,y\rpar\,\lpar 1 + \frac{y}{T}\rpar\Bigr]^{-1/2}\!
= 2\,\pi\,\bigl[ t (1-t)\bigr]^{-1/2}\,J(t),
$$
\vspace{-0.3cm}
\bqa
J(t) &=& \frac{1}{2 \pi i}\,\int_{-i\,\infty}^{+i\,\infty}\,ds\,
B\lpar s ,\frac{1}{2}-s\rpar\,\lpar - \beta^2 - i\,0\rpar^{s-1/2}\,
\intfx{y}\,y^{-1/2}\,(1-y)^{-2\,s}\,\lpar 1+T\,y\rpar^{s-1/2}\,
\lpar 1 + \frac{y}{T}\rpar^{s-1/2}
\nl
{}&=& \frac{1}{2 \pi i}\int_{-i\,\infty}^{+i\,\infty}\!\!\!\!\!ds\,
\frac{\egam{s}\egam{1/2-s}\egam{1-2 s}}{\egam{3/2-2 s}}
\lpar - \beta^2 - i\,0\rpar^{s-1/2}
F_1\lpar \frac{1}{2},\frac{1}{2}-s,
\frac{1}{2}-s,\frac{3}{2}-2s;- T,- \frac{1}{T}\rpar,
\eqa
$0 < \Reb s < 1/2$. Here $F_1$ denotes the first Appell-function.
To obtain the expansion corresponding to $\beta \to 0$ we close the 
integration contour over the right-hand complex half-plane at 
infinity. The leading (double) pole is at $s = 1/2$. Therefore, we obtain
\bq
J(t) = -\,\frac{1}{2}\,\ln \lpar - \beta^2 - i\,0 \rpar 
     + {\cal O}(1),\quad \beta\to0.
\eq
Inserting it into \eqn{VKsing} and using 
$\intfx{t}\,t^{-1/2}\,\lpar 1 - t\rpar^{-3/2}\,\ln t = -\,2\,\pi$,
we get 
\bq
V^{\ssK}_{\rm sing} = \frac{4\,\pi^2}{s^2}\,\ln \lpar - \beta^2 - i\,0 \rpar 
     + {\cal O}(1),\quad \beta\to0.
\label{ourR}
\eq
If the massive loop in \fig{TLvertbca} is made of top quarks the
contribution of $V^{\ssK}$ to the amplitude behaves like
$\beta^2\,V^{\ssK}$ and, therefore, the logarithmic singularity is
$\beta^2\,$-protected at threshold; the same is not true for a
$W\,$-loop. Our result, \eqn{ourR}, is confirmed by the evaluation of
$V^{\ssK}$ of Ref.~\cite{Anastasiou:2006hc} in terms of generalized log
- sine functions. Starting from Eq.(6.34) of
Ref.~\cite{Anastasiou:2006hc} and using the results of
Ref.~\cite{Kalmykov:2004xg} we expand around $\theta = \pi$, where $x =
e^{i\,\theta} = (\beta - 1)\,(\beta + 1)$, with $0 < \theta < \pi$.
This gives for the leading behavior of $V^{\ssK}$ below threshold  
$(\pi^2/2)\,\ln(\theta - \pi)$, where $\ln (-\beta^2) = \ln (\theta-\pi)^2 -
\ln 2$. The same behavior can also be extracted from the results of 
Ref.~\cite{Harlander:2005rq}.
\subsection{Complex masses}
\label{subsec:complexmass}
Our pragmatical solution to the problems induced by unstable particles has been
to remove the $\Reb$ label in those terms that, coming from finite 
renormalization, give $\Reb\,W$ in the Ward identity of \eqn{vWI}.
Furthermore, we decompose \eqn{tfinR} according to:
\bq
{\cal A}_{\rm phys} = 
\Bigl(\sqrt{2}\,\gf\,\mws\Bigr)^{1/2}\frac{\alpha}{2\,\pi}\,A_{\rm
  phys},
\quad
A_{\rm phys} = 
  A^{(1)}_{\,\rm ex}
+ \,\frac{\gf\,\mws}{2\,\sqrt{2}\,\pi^2}\,
  \biggl[
    \frac{A^{(2)}_{\ssR}}{\beta}
  + A^{(2)}_{\ssL}\,\ln\lpar - \beta^2 - i\,0\rpar
  + A^{(2)}_{\rm reg}
  \biggr].
\label{decompo}
\eq
and prove that, as expected, $A^{(2)}_{\ssR}$, $A^{(2)}_{\ssL}$ and 
$A^{(2)}_{\rm reg}$ satisfy (separately) the Ward identity. The latter fact 
allows us to -- minimally -- modify $A^{(2)}_{\ssR,\ssL}$ by working in the 
complex-mass scheme of Ref.~\cite{Denner:2005fg}, i.e. we include complex
masses in the, gauge-invariant, leading part of the two-loop amplitude as
well as in the one-loop part. 

The decomposition of \eqn{decompo} deserves a further comment. There are three
sources of $1/\beta\,$-terms: a) pure two-loop diagrams of the 
$V_{\ssM}\,$-family, i.e. bubble insertions on the internal lines of the 
one-loop triangle; b) $W$-mass renormalization, i.e. on-shell $W$-self-energy 
$\,\times\,$ the mass squared derivative of the one-loop $W\,$-triangle (the
latter giving rise to $1/\beta$); c) Higgs wave-function renormalization
$\,\times\,$ lowest order (the former giving rise to $1/\beta$).

One can easily prove that only c) survives and a,b) that are separately
singular add up to a finite contribution ($\beta \to 0$); their divergency
is an artifact of expanding Dyson resummed propagators.

The $\ln\beta\,$-term originates from pure two-loop diagrams (the 
$V_{\ssK}\,$-family) and it is a remnant of the one-loop Coulomb singularity
of one-loop sub-diagrams.
\section{Numerical results}
The partial width of the Higgs-boson decay into two photons can be
written as
\bq
\Gamma(H\rightarrow\gamma\gamma)=
\frac{\alpha^2\,\gf\,\mw^2}{32\,\sqrt{2}\,\pi^3\,\mh}\left|A_{\rm phys}\right|^2\,.
\eq
The relative correction $\delta$ induced by two-loop (NLO) effects is 
given by $\Gamma=\Gamma_{0}(1+\delta)$, where $\Gamma_0$ is the
lowest order result. It can be split into electroweak and QCD
contributions, $\delta = \delta_{\EW}+\delta_{\QCD}$.

For the numerical evaluation we use the following set of parameters:
\[
\begin{array}{llll}
\mw = 80.398\,\GeV,  \;\; & \;\; 
\mz = 91.1876\,\GeV, \;\; & \;\;
\mt = 170.9\,\GeV,   \;\; & \;\;
\Gamma_{\ssW} = 2.093\,\GeV,\\
\gf = 1.16637\,\times\,10^{-5}\,\GeV^{-2},  \;\; & \;\; 
\alpha(0) = 1/137.0359911,                  \;\; & \;\; 
\alpha_{\ssS}\lpar \mz\rpar= 0.118.         \;\; & \;\; 
\end{array}
\]
All light fermion masses are set to zero and we define the $W\,$-boson
complex pole~\cite{Argyres:1995ym} by
\bq
s_{\ssW} = \mu_{\ssW}\,\lpar \mu_{\ssW} - i\,\gamma_{\ssW}\rpar,
\quad 
\mu^2_{\ssW} = \mws - \Gamma^2_{\ssW},
\quad
\gamma_{\ssW} = \Gamma_{\ssW}\,\lpar 1 - \tfrac{\Gamma^2_{\ssW}}{2\,\mws}\rpar.
\label{replacement}
\eq
\begin{figure}[htb]
\includegraphics[height=6.cm,width=7.cm]{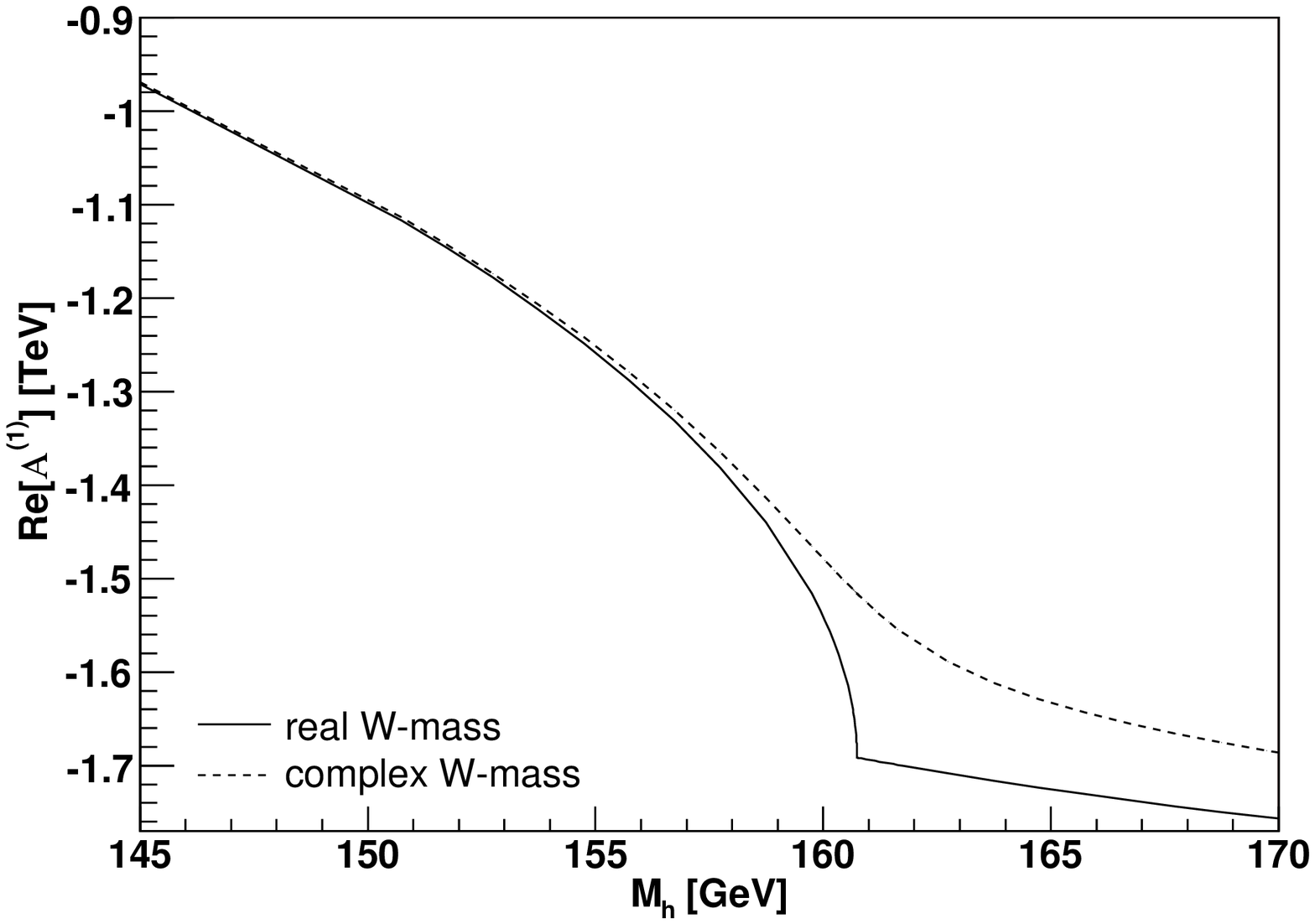}
\hspace{1.cm}
\includegraphics[height=6.cm,width=7.cm]{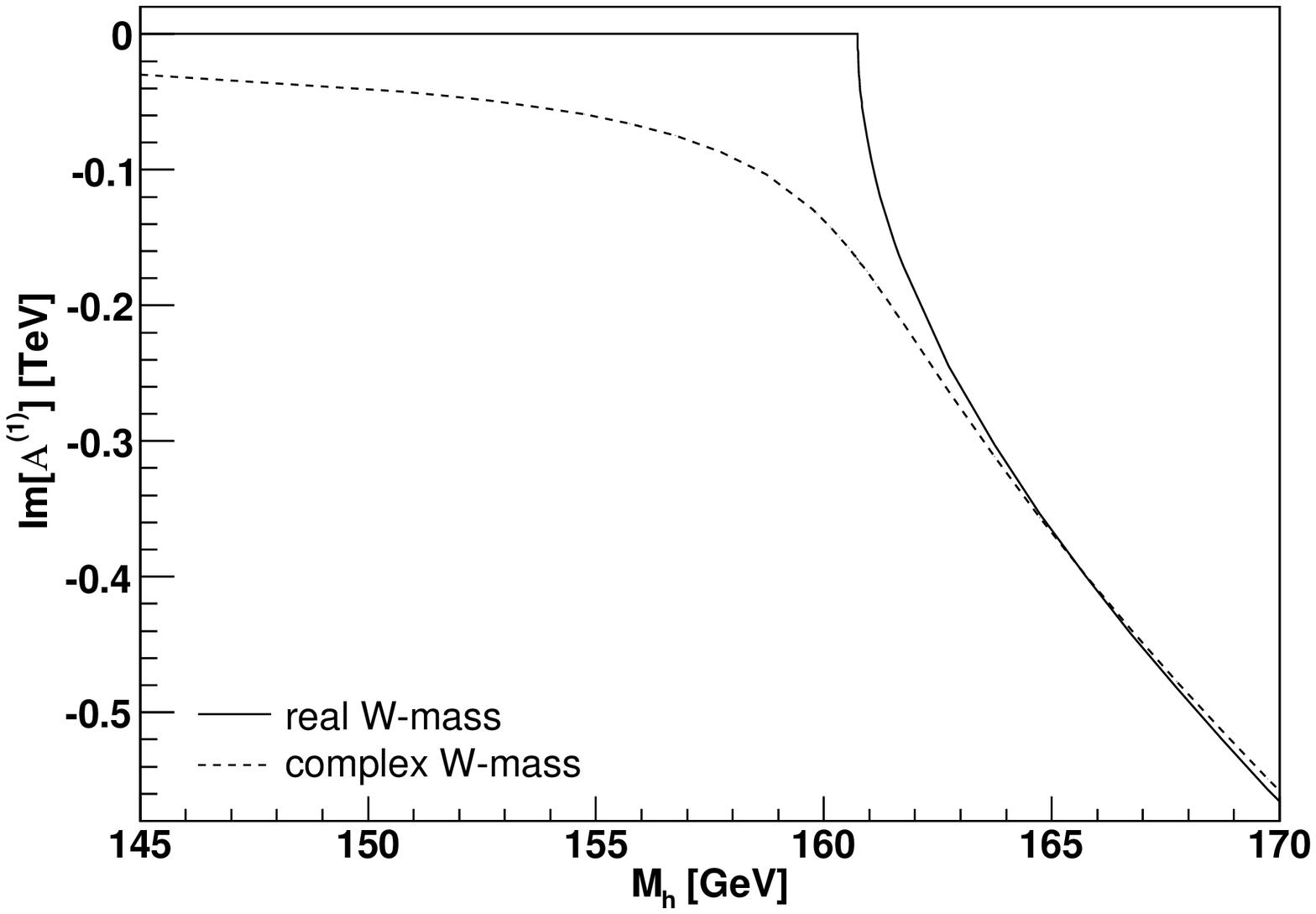}
\vspace{-0.4cm}
\caption[]{\label{OLFigure} Real and imaginary part of the one-loop $H
\to \gamma \gamma$ amplitude with real and complex $W\,$-boson mass.}
\end{figure}
The one-loop $H \to \gamma \gamma$ amplitude, with a complex $W\,$-mass, is 
shown in \fig{OLFigure} around the $WW\,$-threshold including a comparison
with the real $W\,$-mass amplitude.

\begin{figure}[htb]
\begin{center}
\includegraphics[height=7.cm,width=9.cm]{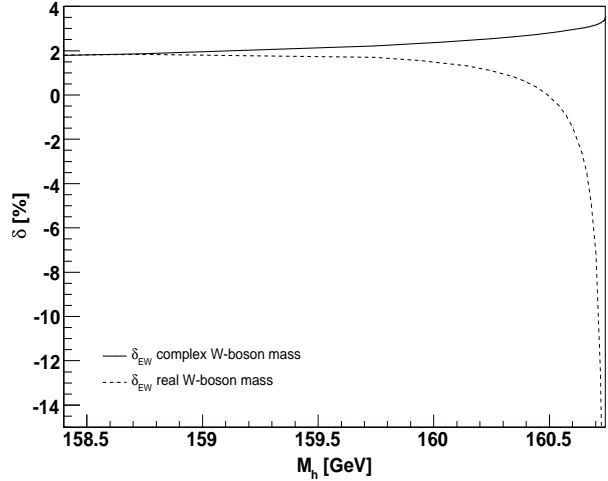}
\end{center}
\vspace{-0.6cm}
\caption[]{\label{CMassFigure} Percentage electroweak corrections to $H
\to \gamma \gamma$ with real (dashed) and complex (solid) $W\,$-boson mass,
below the $WW\,$-threshold.}
\end{figure}

A comparison of the percentage electroweak corrections, with and without 
complex $W$-masses, is shown in \fig{CMassFigure} for a Higgs mass range below 
the $WW\,$-threshold showing the unphysical growth of the real case and also
some sizable difference in a region of about two GeV below the threshold.

\begin{figure}[!ht]
\includegraphics[height=6.cm,width=7.cm]{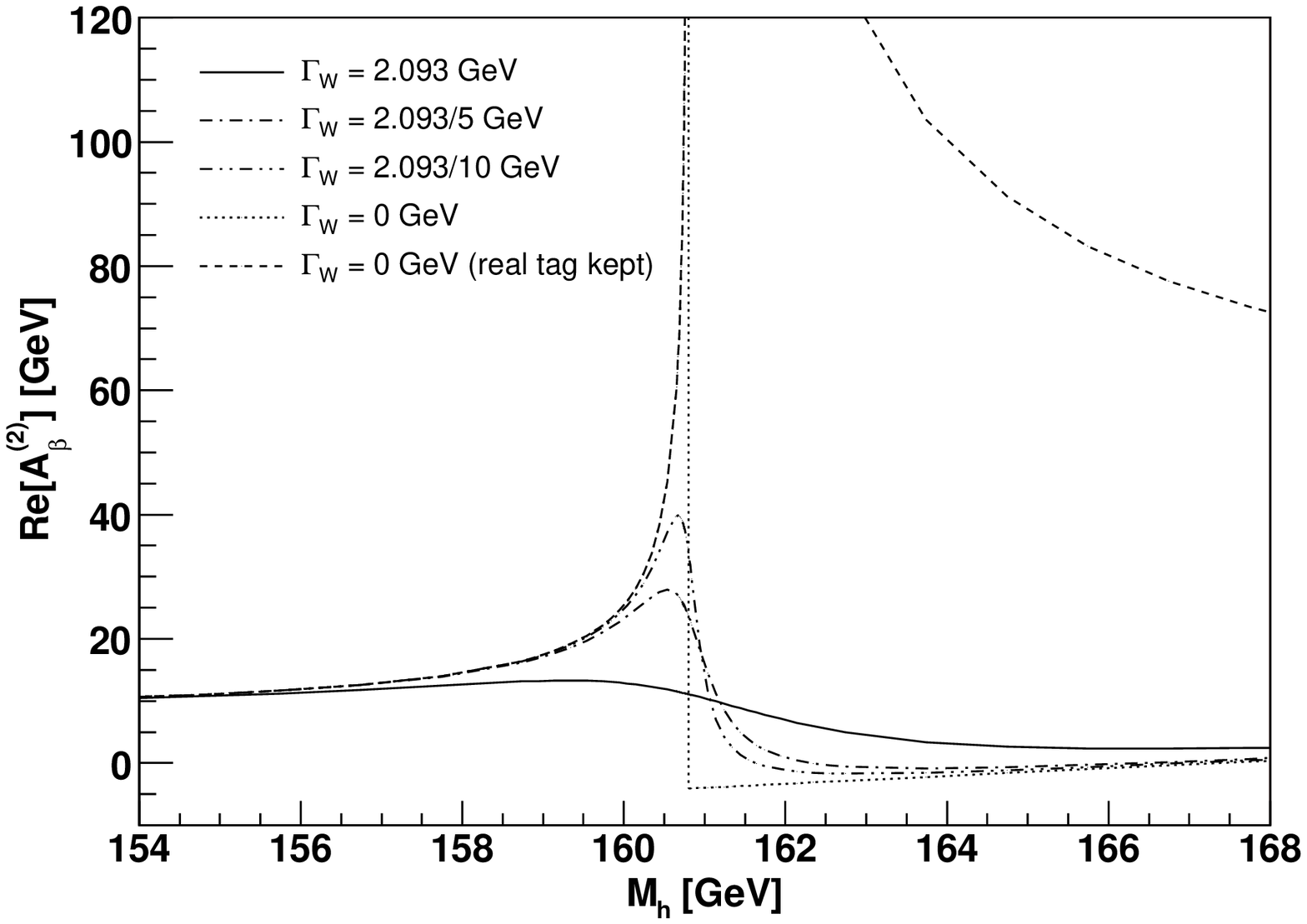}
\hspace{1.cm}
\includegraphics[height=6.cm,width=7.cm]{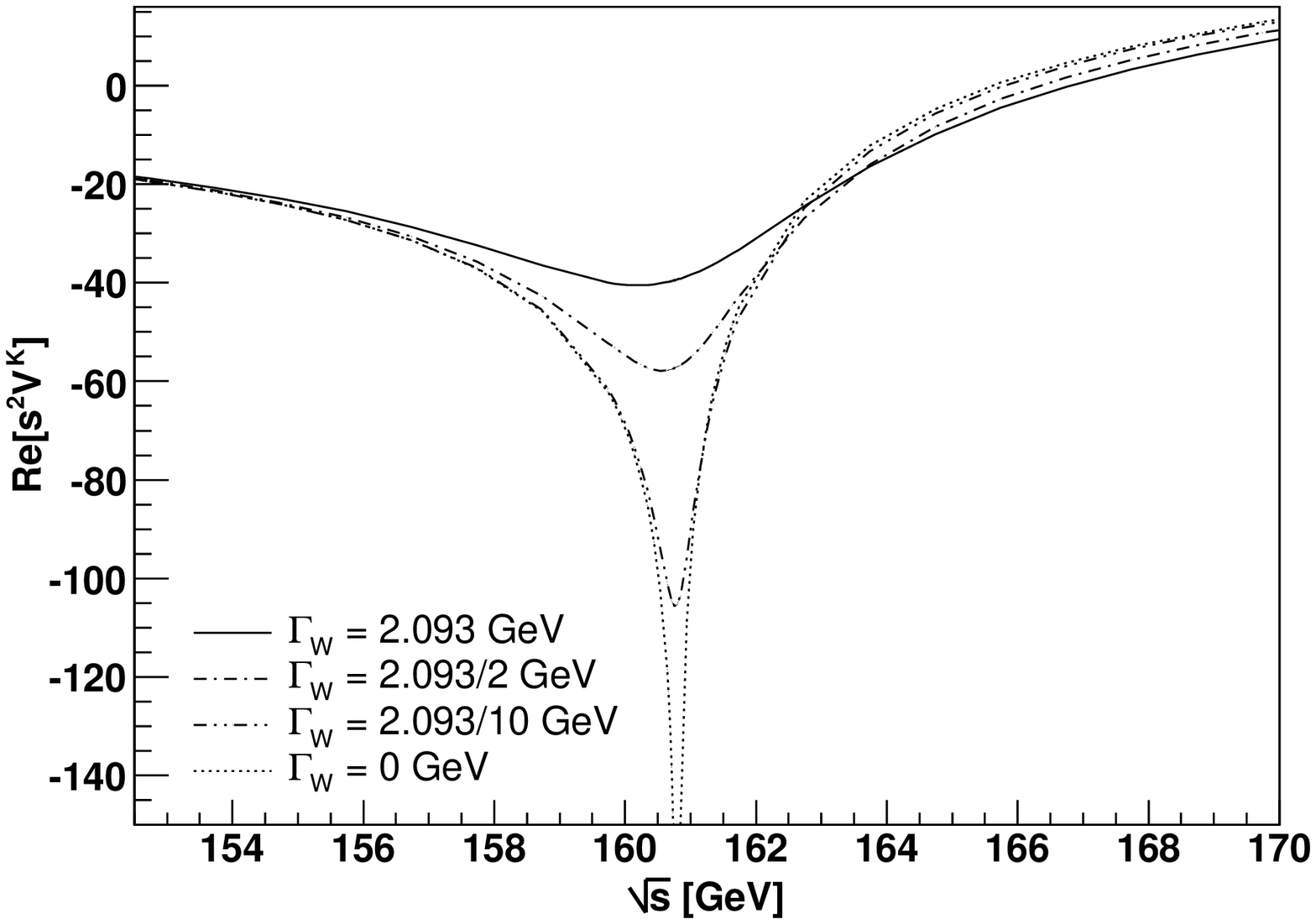}
\vspace{-0.4cm}
\caption[]{\label{VGFigure} The effect of varying the $W\,$-boson width
in the real part of
$A^{(2)}_{\beta}=\tfrac{\gf\,\mws}{2\,\sqrt{2}\,\pi^2}\,\tfrac{A_{R}^{(2)}}{\beta}$
containing the $1/\beta$ terms of the two-loop amplitude (left) and in the
real part of $V^{\ssK}$ of \fig{TLvertbca} containing the
$\ln\!\beta$ terms (right) is shown. We also show the effect of using a
real $W\,$-boson mass but removing $\Reb\,$-labels in finite
renormalization.
}
\end{figure}

We have also analyzed the effect of (artificially) varying the imaginary part
of the $W\,$-boson complex mass; results are given in \fig{VGFigure},
showing that our {\em complex} result reproduces the {\em real} one in the
limit $\Gamma_{\ssW} \to 0$. \fig{VGFigure} clearly demonstrates the large
but artificial effects arising at normal-thresholds of unstable particles when
their masses are kept real.

\begin{figure}[!ht]
\begin{center}
\includegraphics[width=14.cm]{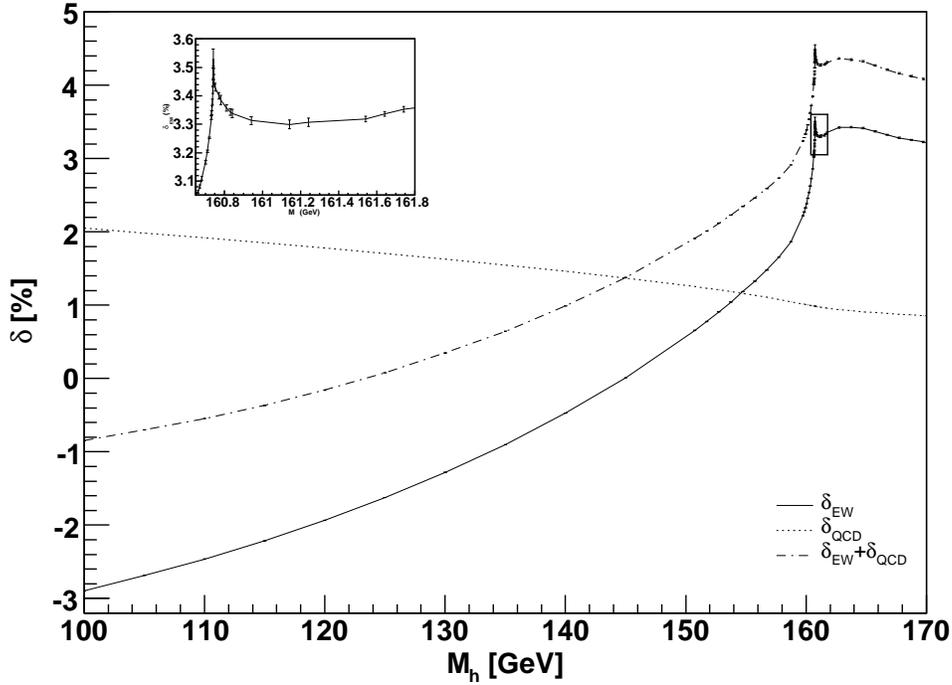}
\end{center}
\vspace{-0.7cm}
\caption[]{\label{DeltaFigure} Electroweak (solid), QCD (dashed) and
  total (dashed-dotted) correction in percent for the partial width of
  the decay $H \to \gamma \gamma$.}
\end{figure}

Finally, in \fig{DeltaFigure} we show both QCD and electroweak percentage 
corrections to the decay width $\Gamma(H \to \gamma \gamma)$, including the region 
around the $WW\,$-threshold. A running $\alpha_s$ has been used for the 
computation of the QCD corrections. 
The remaining cusp of $\delta_{\EW}$ at the $WW\,$-threshold, whose
details are shown in the blow-up of \fig{DeltaFigure}, is due to our 
{\em minimal} scheme where the $W\,$-mass is kept real in 
$A^{(2)}_{\rm reg}$, the regular part of the amplitude (see \eqn{decompo}).
The relatively small error bars in a region so close to threshold serve as
evidence for the efficiency of our numerical algorithms.

Our result for $\delta_{\EW}$ in the region $100\,\GeV < \mh < 150\,\GeV$
is in substantial agreement with those of Ref.~\cite{Degrassi:2005mc}. 
In conclusion, we observe a cancellation of the two corrections below the 
threshold whereas, above it, both $\delta_{\QCD}$ and $\delta_{\EW}$ are 
positive leading to a sizable (up to $4.5\%$) total correction to the decay
width. The perturbative expansion for the decay rate, supplemented with the
complex-mass scheme, gives reliable and accurate predictions in a wide range
of values for the Higgs-boson mass, typically $-1\% < \delta_{\rm tot} < 
4\%$ in the range $100\,\GeV < \mh < 170\,\GeV$.
\section{Conclusions}
In this paper we provide a stand-alone numerical calculation of the full
two-loop corrections to the decay width $\Gamma(H \to \gamma \gamma)$. Since no
expansion is involved in the calculation we can produce results for all values
of the Higgs-boson mass, as shown in \fig{DeltaFigure}, including the 
$WW\,$-threshold. The techniques introduced in this context are general
enough to be used for all kinematical configurations of $1 \to 2$ processes 
at the two-loop level.

To deal with normal-threshold singularities, specifically the
$WW\,$-threshold, we have introduced complex $W\,$-masses in a
gauge-invariant manner; our {\em minimal} scheme selects gauge-invariant
components, typically LO (one-loop) amplitude and divergent parts of the
NLO (two-loop) amplitude, and perform the replacement of
\eqn{replacement}.  Details of our approach will be described in a
forthcoming publication.

The main result obtained in this paper can be summarized by saying that
the NLO percentage corrections to the decay width $\Gamma(H \to \gamma \gamma)$,
$\delta_{\QCD}$ and $\delta_{\EW}$, compensate below threshold leading
to a small total correction; however, above the $WW\,$-threshold they are
both positive, leading to a sizable overall effect of $\approx 4\%$. 
\Acknowledgments
We gratefully acknowledge the invaluable assistance of Stefano Actis in 
several steps of this calculation. 


\end{document}

%% file: sbookp_rosetta.tex
\newcommand{\nl}{\nonumber\\}
\newcommand{\nn}{\nonumber}

\newcommand{\lpar}{\left(}                            
\newcommand{\rpar}{\right)}

\newcommand{\bq}{\begin{equation}}                    
\newcommand{\eq}{\end{equation}}
\newcommand{\bqa}{\arraycolsep 0.14em\begin{eqnarray}}
\newcommand{\eqa}{\end{eqnarray}}
\newcommand{\ba}[1]{\begin{array}{#1}}
\newcommand{\ea}{\end{array}}
\newcommand{\ben}{\begin{enumerate}}
\newcommand{\een}{\end{enumerate}}
\newcommand{\bei}{\begin{itemize}}
\newcommand{\eei}{\end{itemize}}
\newcommand{\eqn}[1]{Eq.(\ref{#1})}

\newcommand{\fig}[1]{Fig.~\ref{#1}}

\newcommand{\GeV}{\mathrm{GeV}}

\def\Re{\mathop{\operator@font Re}\nolimits}
\def\Im{\mathop{\operator@font Im}\nolimits}
\newcommand{\ord}[1]{{\cal O}\lpar#1\rpar}

\newcommand{\mw}{M_{_W}}

\newcommand{\mz}{M_{_Z}}

\newcommand{\mh}{M_{_H}}

\newcommand{\mt}{m_t}

\newcommand{\mws}{M^2_{_W}}

\newcommand{\mzs}{M^2_{_Z}}

\newcommand{\mhs}{M^2_{_H}}

\newcommand{\gf}{G_{\ssF}}

\newcommand{\stws}{s_{\theta}^2}

\newcommand{\ctws}{c_{\theta}^2}

\newcommand{\li}[2]{\mathrm{Li}_{#1}\lpar\displaystyle{#2}\rpar} 

\newcommand{\egam}[1]{\Gamma\lpar#1\rpar}               

\newcommand{\intfx}[1]{\int_{\scriptstyle 0}^{\scriptstyle 1}\,d#1}

\newcommand{\ep}{\epsilon}

\newcommand{\Reb}{{\rm{Re}}}
\newcommand{\Imb}{{\rm{Im}}}
\newcommand{\upar}[1]{u}

\newcommand{\ssA}{{\scriptscriptstyle{A}}}
\newcommand{\ssB}{{\scriptscriptstyle{B}}}
\newcommand{\ssC}{{\scriptscriptstyle{C}}}
\newcommand{\ssD}{{\scriptscriptstyle{D}}}
\newcommand{\ssE}{{\scriptscriptstyle{E}}}
\newcommand{\ssF}{{\scriptscriptstyle{F}}}
\newcommand{\ssG}{{\scriptscriptstyle{G}}}
\newcommand{\ssH}{{\scriptscriptstyle{H}}}

\newcommand{\ssK}{{\scriptscriptstyle{K}}}
\newcommand{\ssL}{{\scriptscriptstyle{L}}}
\newcommand{\ssM}{{\scriptscriptstyle{M}}}

\newcommand{\ssP}{{\scriptscriptstyle{P}}}

\newcommand{\ssR}{{\scriptscriptstyle{R}}}
\newcommand{\ssS}{{\scriptscriptstyle{S}}}

\newcommand{\ssW}{{\scriptscriptstyle{W}}}

\newcommand{\QCD}{\rm{\scriptscriptstyle{QCD}}}
\newcommand{\EW}{\rm{\scriptscriptstyle{EW}}}

\newcommand{\bqas}{\begin{eqnarray*}}
\newcommand{\eqas}{\end{eqnarray*}}

\def\app#1#2 {{\it Acta. Phys. Pol.} {\bf#1},#2}
\def\cpc#1#2 {{\it Computer Phys. Comm.} {\bf#1},#2}
\def\np#1#2 {{\it Nucl. Phys.} {\bf#1},#2}
\def\pl#1#2 {{\it Phys. Lett.} {\bf#1},#2}
\def\prep#1#2 {{\it Phys. Rep.} {\bf#1},#2}
\def\prev#1#2 {{\it Phys. Rev.} {\bf#1},#2}
\def\prl#1#2 {{\it Phys. Rev. Lett.} {\bf#1},#2}
\def\zp#1#2 {{\it Zeit. Phys.} {\bf#1},#2}
\def\sptp#1#2 {{\it Suppl. Prog. Theor. Phys.} {\bf#1},#2}
\def\mpl#1#2 {{\it Modern Phys. Lett.} {\bf#1},#2}
\def\jetp#1#2 {{\it Sov. Phys. JETP} {\bf#1},#2}
\def\fpj#1#2 {{\it Fortschr. Phys.} {\bf#1},#2}
\def\afp#1#2 {{\it Acta.Phys. Polon.} {\bf#1},#2}
\def\err#1#2 {{\it Erratum} {\bf#1},#2}
\def\ijmp#1#2 {{\it Int. J. Mod. Phys} {\bf#1},#2}
\def\nc#1#2 {{\it Nuovo Cimento} {\bf#1},#2}
\def\ap#1#2 {{\it Ann. Phys.} {\bf#1},#2}
\def\cmp#1#2 {{\it Comm. Math. Phys.} {\bf#1},#2}
\def\el#1#2 {{\it Europhys. Lett.} {\bf#1},#2}
\def\hpa#1#2 {{\it Helv. Phys. Acta} {\bf#1},#2}
\def\yf#1#2 {{\it Yad. Fiz.} {\bf#1},#2}
\def\nim#1#2 {{\it Nucl. Instrum. Meth.} {\bf#1},#2}
\def\spz#1#2 {{\it Sov. Pisma Zhetf} {\bf#1},#2}
\def\jetpl#1#2 {{\it JETP Lett.} {\bf#1},#2}
\def\sjnp#1#2 {{\it Sov. J. Nucl. Phys.} {\bf#1},#2}
\def\ptp#1#2 {{\it Progr. Theor. Phys. (Kyoto)} {\bf#1},#2}
\def\rmp#1#2  {{\it Rev. Mod. Phys.} {\bf#1},#2}
\def\zhetf#1#2 {{\it ZhETF} {\bf#1},#2}
\def\prs#1#2 {{\it Proc. Roy. Soc.} {\bf#1},#2}
\def\phys#1#2 {{\it Physica} {\bf#1},#2}

\def\bfi{\begin{figure}}
\def\efi{\end{figure}}

\newcommand{\GS}{{\cal G}raph{\cal S}hot}

\newcommand{\intsx}[1]{\int_{\scriptstyle 0}^{\scriptstyle 1}\!\!\!d#1}

\newcommand{\bmid}{\Bigr|}

\newcommand{\ssWW}{{\scriptscriptstyle{W}\scriptscriptstyle{W}}}
